\documentclass[10pt,conference]{IEEEtran}
\usepackage{epsfig,setspace,amsmath,epsf,amssymb,bm,theorem,cite,graphicx, epstopdf,algorithm,algpseudocode,float,color,mathtools}
\usepackage[table,xcdraw]{xcolor}

\newtheorem{theorem}{Theorem}

\newtheorem{remark}{Remark}

\newcommand{\bS}{{\bm{S}}}
\newcommand{\be}{{\bm{e}}}

\IEEEoverridecommandlockouts
\allowdisplaybreaks

\begin{document}

\title{Private Federated Submodel Learning \\ with Sparsification}

\author{Sajani Vithana \qquad Sennur Ulukus\\
	\normalsize Department of Electrical and Computer Engineering\\
	\normalsize University of Maryland, College Park, MD 20742 \\
	\normalsize \emph{spallego@umd.edu} \qquad \emph{ulukus@umd.edu}}

\maketitle

\begin{abstract}
We investigate the problem of private read update write (PRUW) in federated submodel learning (FSL) with sparsification. In FSL, a machine learning model is divided into multiple submodels, where each user updates only the submodel that is relevant to the user's local data. PRUW is the process of privately performing FSL by reading from and writing to the required submodel without revealing the submodel index or the values of updates to the databases. Sparsification is a widely used concept in learning, where the users update only a small fraction of parameters to reduce the communication cost. Revealing the coordinates of these selected (sparse) updates leaks privacy of the user. We show how PRUW in FSL can be performed with sparsification. We propose a novel scheme which privately reads from and writes to arbitrary parameters of any given submodel, without revealing the submodel index, values of the updates, or the coordinates of the sparse updates, to databases. The proposed scheme achieves significantly lower reading and writing costs compared to what is achieved without sparsification.
\end{abstract}

\section{Introduction}

Federated learning (FL) \cite{FL1,FL2,Advances,magazine} enables distributed machine learning without the users having to share their private data directly with a central server. This solves user privacy concerns to a certain extent and decentralizes the processing power requirements. However, the communication cost of FL is considerable as a large number of users keep sharing gradient updates with the central server in an iterative manner. Gradient sparsification \cite{sparse1,GGS,rtopk,adaptive,conv,overtheair,timecorr,qsl,conv2}, gradient quantization \cite{qsgd,fedpaq,constraints}, and federated submodel learning (FSL) \cite{rw_jafar,ourICC,dropout,pruw,paper1,billion,secureFSL} are some of the solutions proposed for this problem. In this work, we propose a novel scheme that combines FSL and gradient sparsification while preserving information-theoretic privacy of the users. 

In FSL, a machine learning model is divided into multiple submodels based on different types of data used to train parts of the model. In FSL, a given user downloads (reads) and updates (writes) only the submodel that is relevant to the user's local data, which reduces the communication cost. In this process, the updated submodel index as well as the values of the updates leak information about the type of data that the user has. Thus, in order to perform FSL while preserving user privacy, both the submodel index and the values of the updates need to be kept private in both reading and writing phases. This is known as private read update write (PRUW)  \cite{rw_jafar,ourICC,dropout,pruw,paper1,billion,secureFSL}. The reading phase of PRUW is similar to private information retrieval (PIR) \cite{original, PIR, coded, chaoTian, leaky, SPIR, PSI, colluding, MMPIR, byzantine, semanticPIR, singleDB, XSTPIR, sideinfo, Kumar_PIRarbCoded}.

Gradient sparsification is a widely used technique in FL, where users send only a small fraction of updates (gradients) to the servers in order to reduce the communication cost. In this case, the users send only the updated values (randomly chosen or largest amplitude values) along with their positions to the servers. However, directly sending the updates and their positions leaks user's privacy \cite{comprehensive,featureLeakage,InvertingGradients,DeepLeakage,BeyondClassRepresentatives,recent,MembershipInterference,SecretSharer,PracticalSecureAgg,reinforcement,avgDP,cpSGD,PrivacyAmp,DPFL,shuffle,PrivacyBlanket,shuffledDPFL,client}. In this paper, we propose a method to privately send the subset of updates along with their indices using a noisy shuffling mechanism.

\begin{figure}[t]
    \centering
    \includegraphics[scale=0.5]{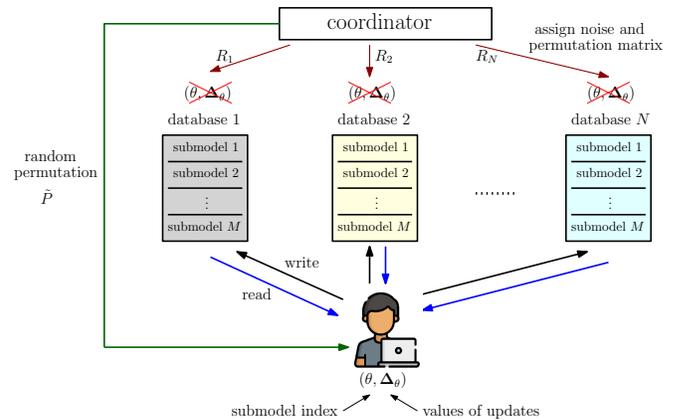}
     \vspace*{-0.2cm}
    \caption{System model.}
    \label{fig1}
    \vspace*{-0.5cm}
\end{figure}

PRUW in FSL with sparsification consists of two phases, the reading phase in which a user privately downloads the required submodel, and the writing phase in which the user privately uploads the updates back to the databases. Each user engages in the PRUW process in an iterative manner. In the writing phase, each user only uploads a small fraction of updates to reduce the writing cost. In the reading phase, each user only downloads the parameters that were updated in the previous iteration. In this process, the updating submodel index and the values of the updates are kept private. Thus, we need to ensure that the actual update/position pairs of the sparse updates (even with added noise) are not directly sent to the databases since they reveal the values of those updates whose indices were not specified. As a solution to this problem, we propose a method which is an extension of the PRUW scheme in \cite{ourICC}. The system model, as in \cite{ourICC} and \cite{dropout}, includes a coordinator that helps initialize the PRUW process. The coordinator in \cite{ourICC,dropout} is necessary to initialize the same noise terms in the storage across the $N$ databases. In this work, the coordinator is used again to introduce a shuffling mechanism that hides the real positions of the sparse updates from databases, which guarantees privacy in FSL with sparsification.

The major contributions of this work are: 1) introduction of the concept of sparsification in FSL to achieve significantly lower communication costs while still achieving information-theoretic privacy, 2) a novel scheme that performs PRUW with arbitrary sparsification rates with zero information leakage. 

\section{Problem Setting}

Consider $M$ independent submodels, each having $P$ subpackets, stored in $N$ non-colluding databases. At a given time instance $t$, a user reads, updates and writes to one of the $M$ submodels, while not revealing any information about the updated submodel index or the values of updates. The storage, queries and updates consist of symbols from a finite field $\mathbf{F}_q$.

In the PRUW process, users keep reading, updating and writing to required submodels in an iterative manner. A given user at time $t$ first downloads (reads) all subpackets of the required submodel privately. However, based on the concept of sparsification in learning, the user only writes to an $r$ fraction of subpackets of the required submodel.\footnote{In the update stage (model training) users typically work in continuous fields (real numbers) and make $1-r$ of the updates equal to zero (i.e., not update) based on the concept of sparsification in learning. These updates are then converted to symbols in $\mathbf{F}_q$, to be sent to the databases. We assume that the zeros in the continuous field are converted to zeros in the finite field.} This drastically reduces the writing cost. Therefore, a given user who reads the same submodel at time $t+1$ only has to download the union of each $r$ fractions of subpackets updated by all users at time $t$. Let the cardinality of this union be $Pr'$, where $0\leq r'\leq1$.

The reduction in the communication cost of the PRUW process with sparsification results from communicating the non-zero-valued updates and their positions to the databases (users) in the writing (reading) phase. However, this leaks information about $1-r$ of the updates in the writing phase, as their values (zero) are revealed to the databases. In order to avoid this, we use a shuffling method, in which actual positions of the non-zero updates are randomly shuffled with the aid of a coordinator as shown in Fig.~\ref{fig1}. This ensures that the indices of zero-valued parameters are hidden from the databases. 

The three components in PRUW with sparsification that need to be kept private are: 1) index of the submodel updated by each user, 2) values of the updates, and 3) indices (positions) of the sparse updates. Note that 3 is a requirement that is implied by 2. The formal descriptions of the privacy constraints are given below. The constraints are presented in the perspective of a single user at time $t$, even though multiple users update the model simultaneously.

\emph{Privacy of the submodel index:} No information on the index of the submodel that is being updated, $\theta$, is allowed to leak to any of the databases, i.e., for each $n$, 
\begin{align}
    I(\theta^{[t]};Q_n^{[t]},Y_n^{[t]}|Q_n^{[1:t-1]},Y_n^{[1:t-1]},S_n^{[1:t-1]})=0,
\end{align}
where $Q_n$ and $Y_n$ are the queries and updates/coordinates sent by a given user to database $n$ in the reading and writing phases, at corresponding time instances indicated in square brackets and $S_n$ is the storage of database $n$.

\emph{Privacy of the values of updates:} No information on the values of the updates is allowed to leak to any of the databases, i.e., for each $\tilde{q}\in\mathbf{F}_q$,
\begin{align}
    P(\Delta_{\theta,i}^{[t]}=\tilde{q}|Q_n^{[1:t]},Y_n^{[1:t]})=P(\Delta=\tilde{q}), 
\end{align}
for each database $n$, where $\Delta_{\theta,i}^{[t]}$ is the update of the $i$th parameter of submodel $\theta$ generated by a given user at time $t$. $P(\Delta=\tilde{q})$, $\tilde{q}\in\mathbf{F}_q$ is the globally known apriori distribution of any given parameter update characterized by,\footnote{We assume that all parameters in each of the most significant $r$ fraction of subpackets have non-zero updates.}
\begin{align}
    P(\Delta=\tilde{q})=\begin{cases}
    1-r, & \text{if $\tilde{q}=0$},\\
    \frac{r}{q-1}, & \text{for each $\tilde{q}\neq 0$}.
    \end{cases}
\end{align}

\emph{Security of submodels:} No information on the submodels is allowed to leak to any of the databases, i.e., for each $n$,
\begin{align}
    I(W_{1:M}^{[t]};S_n^{[t]})=0,
\end{align}
where $W_k^{[t]}$ is the $k$th submodel at time $t$.

\emph{Correctness in the reading phase:} The user should be able to correctly decode the required submodel from the answers received in the reading phase, i.e., 
\begin{align}
H(W_{\theta}^{[t-1]}|Q_{1:N}^{[t]},A_{1:N}^{[t]})=0,
\end{align}
where $A_n^{[t]}$ is the answer from database $n$ at time $t$.

\emph{Correctness in the writing phase:} Each parameter $i$ of the non-zero $Pr$ subpackets of $W_{\theta}$ must be correctly updated at time $t$ as (for a single user),
\begin{align}
    W_{\theta,i}^{[t]}=W_{\theta,i}^{[t-1]}+\Delta_{\theta,i}.
\end{align}

\emph{Reading and writing costs:} The reading and writing costs are defined as $C_R=\frac{\mathcal{D}}{L}$ and $C_W=\frac{\mathcal{U}}{L}$, respectively, where $\mathcal{D}$ is the total number of bits downloaded in the reading phase, $\mathcal{U}$ is the total number of bits uploaded in the writing phase, and $L$ is the size of each submodel. The total cost $C_T$ is the sum of the reading and writing costs, $C_T=C_R+C_W$.

\section{Main Result}

\begin{theorem}
In a private FSL setting with $N$ databases, $M$ submodels, $P$ subpackets in each submodel, and $r$ and $r'$ sparsification rates in the uplink and downlink, respectively, the following reading and writing costs are achievable,
\begin{align}
    C_R&=\frac{4r'+\frac{4}{N}(1+r')\log_qP}{1-\frac{2}{N}},\label{re}\\
    C_W&=\frac{4r(1+\log_q P)}{1-\frac{2}{N}}.\label{wr}
\end{align}

\end{theorem}

\begin{remark}
If sparsification is not considered in the PRUW process, the lowest achievable reading and writing costs are given by $C_R=C_W=\frac{2}{1-\frac{2}{N}}$; see \cite{ourICC}. Therefore, sparsification with smaller values of $r$ and $r'$ results in significantly reduced communication costs as shown in \eqref{re} and \eqref{wr}.
\end{remark}

\begin{remark}
The proposed PRUW scheme with sparsification defined for private FSL can also be modified and applied to private FL. The reading and writing costs for this case are given by $C_R=\frac{3r'+\frac{3}{N}(1+r')\log_q P}{1-\frac{1}{N}}$  and $C_W=\frac{3r(1+\log_q P)}{1-\frac{1}{N}}$, respectively. The reduction in the communication cost is at the expense of a larger permutation-reversing matrix.
\end{remark}

\section{Proposed Scheme}

The scheme is similar to what is presented in \cite{ourICC} with the additional component of sparse uploads and downloads. In the writing (reading) phase of the scheme in \cite{ourICC}, the updates (values) of all parameters in a given subpacket are combined into a single bit. Thus, a user sends (receives) $P$ bits per database, where $P$ is the number of subpackets in a submodel. In this work, we assume that the user only updates $Pr$ of the subpackets due to sparsification. Hence, the user only has to send $Pr\ll P$ single bit updates, which reduces the writing cost. Similarly, the user only needs to download the subset of parameters that were updated in the previous iteration. However, revealing the indices of the subpackets with no update (all zeros) in the writing phase leaks privacy, as the values of those updates (zero) are directly known by the databases. Thus, we use random permutations to hide the subpacket indices which have zero updates.

\subsection{Storage and Initialization}

The storage of a single subpacket in database $n$ is,
\begin{align}
    \bS_n=\begin{bmatrix}\label{storage3}
    \begin{bmatrix}
        W_{1,1}+ (f_1-\alpha_n)\sum_{i=0}^{2\ell}\alpha_n^i Z_{1,i}^{[1]}\\
        W_{2,1}+ (f_1-\alpha_n)\sum_{i=0}^{2\ell}\alpha_n^i Z_{2,i}^{[1]}\\
        \vdots\\
        W_{M,1}+ (f_1-\alpha_n)\sum_{i=0}^{2\ell}\alpha_n^i Z_{M,i}^{[1]}\\
    \end{bmatrix}\\
    \vdots\\
    \begin{bmatrix}
        W_{1,\ell}+ (f_\ell-\alpha_n)\sum_{i=0}^{2\ell}\alpha_n^i Z_{1,i}^{[\ell]}\\
        W_{2,\ell}+ (f_\ell-\alpha_n)\sum_{i=0}^{2\ell}\alpha_n^i Z_{2,i}^{[\ell]}\\
        \vdots\\
        W_{M,\ell} + (f_\ell-\alpha_n)\sum_{i=0}^{2\ell}\alpha_n^i Z_{M,i}^{[\ell]}\\
    \end{bmatrix}
    \end{bmatrix}, 
\end{align}
where the subpacketization $\ell=\frac{N-2}{4}$, $W_{i,j}$ is the $j$th bit of the given subpacket of submodel $i$, $W_i$, $Z_{i,j}^{[k]}$ is the $(j+1)$st noise term for the $k$th bit of $W_i$, and $\{f_i\}_{i=1}^\ell$, $\{\alpha_n\}_{n=1}^N$ are globally known distinct constants chosen from $\mathbf{F}_q$, such that each $\alpha_n$ and $f_i-\alpha_n$ for all $i\in\{1,\dotsc,\ell\}$ and $n\in\{1,\dotsc,N\}$ are coprime with $q$. The number of subpackets is $P=\frac{4L}{N-2}$.

In PRUW, at time $t=0$, it should be ensured that all noise terms in storage are the same in all databases. This is handled by the coordinator in Fig.~\ref{fig1}. We make use of this coordinator again in PRUW with sparsification as follows. In the reading and writing phases of PRUW with sparsification, the user only reads and writes parameters/updates corresponding to a subset of subpackets ($\ll P$) without revealing their true indices. The coordinator is used to privately shuffle the true non-zero subpacket indices as explained next.

At the beginning of the FSL system design, $t=0$, the coordinator picks a random permutation of indices $\{1,\dotsc,P\}$ out of all $P!$ options, denoted by $\tilde{P}$, where $P$ is the number of subpackets. The coordinator sends $\tilde{P}$ to all users involved in the PRUW process. Then, the coordinator sends the corresponding noise added permutation-reversing matrix to database $n$, $n\in\{1,\dotsc,N\}$, given by,
\begin{align}\label{rearrange}
    R_n=R+\prod_{i=1}^\ell (f_i-\alpha_n)\bar{Z},
\end{align}
where $R$ is the permulation-reversing matrix and $\bar{Z}$ is a random noise matrix, both of size $P\times P$. For each database, $R_n$ is a random noise matrix, from which nothing can be learnt about the random permutation. The matrix $R_n$ is fixed at database $n$ at all time instances.

\subsection{Reading Phase at Time $t$} \label{readingphase}

The process of reading (downlink) a subset of parameters of a given submodel without revealing the submodel index or the parameter indices within the submodel is explained in this section. Since the PRUW process is carried out in an iterative manner, a given user at time $t$ must only download the subpackets that were updated in the writing phase at time $t-1$. In the writing phase at time $t-1$, multiple individual users update $r$ fractions of subpackets of the relevant submodels. Let $J_i$ be the set of permuted indices of the $Pr$ of subpackets updated by user $i$ for $i=1,\dotsc,\mathcal{U}$, in a case where $\mathcal{U}$ users independently updated the model at time $t-1$. Since each database has access to all $J_i$s, each database calculates $\tilde{V}=\cup_{i} J_i$. One designated database sends $\tilde{V}$ to each user at time $t$. Once the users know the subset of permuted subpacket indices which have been updated at time $t-1$, the users can find the real subpacket indices (permutation-reversed) of the updates, since they know the permutation from the coordinator $\tilde{P}$. The next steps of the reading phase at time $t$ are as follows:
\begin{enumerate}
    \item The user sends a query to each database $n$, to privately specify the required submodel $W_{\theta}$ given by,
    \begin{align}\label{query}
    Q_n=\begin{bmatrix}
        \frac{1}{f_1-\alpha_n}\be_M(\theta)+\Tilde{Z}_{1}\\
        \frac{1}{f_2-\alpha_n}\be_M(\theta)+\Tilde{Z}_{2}\\
        \vdots\\
        \frac{1}{f_\ell-\alpha_n}\be_M(\theta)+\Tilde{Z}_{\ell}
    \end{bmatrix}, \quad n\in\{1,\dotsc,N\},
    \end{align}
    where $\be_M(\theta)$ is the all zeros vector of size $M\times1$ with a $1$ at the $\theta$th position and $\Tilde{Z}_{i}$ are random noise vectors.
    
    \item In order to download the non-permuted version of the $i$th, $i\in\{1,\dotsc,|\tilde{V}|\}$, required subpacket (i.e., $V(i)=\tilde{P}(\tilde{V}(i))$) from the set $\tilde{V}$, database $n$ picks the column $\tilde{V}(i)$ of the permutation-reversing matrix $R_n$ given in \eqref{rearrange} indicated by $R_n(:,\tilde{V}(i))$ and calculates the corresponding query given by,
    \begin{align}
        Q_n^{[V(i)]}&=\begin{bmatrix}
            R_n(1,\tilde{V}(i))Q_n\\
            \vdots\\
            R_n(P,\tilde{V}(i))Q_n
        \end{bmatrix}\\
        &=\begin{bmatrix}\!
            1_{\{V(i)=1\}}\!\!\begin{bmatrix}
                \frac{1}{f_1-\alpha_n}\be_M(\theta)\\
                \vdots\\
                \frac{1}{f_{\ell}-\alpha_n}\be_M(\theta)\\
            \end{bmatrix}\!\!+\!P_{\alpha_n}(\ell)\\
            \vdots\\
            \!1_{\{V(i)=P\}}\!\!\begin{bmatrix}
                \frac{1}{f_1-\alpha_n}\be_M(\theta)\\
                \vdots\\
                \frac{1}{f_{\ell}-\alpha_n}\be_M(\theta)\\
            \end{bmatrix}\!\!+\!P_{\alpha_n}(\ell)
        \end{bmatrix},
    \end{align}
    where $P_{\alpha_n}(\ell)$ are noise vectors consisting of polynomials of $\alpha_n$ of degree $\ell$.
    \item Then, the user downloads subpacket $V(i)=\tilde{P}(\tilde{V}(i))$, $i\in\{1,\dotsc,|\tilde{V}|\}$ of the required submodel using the answers received by the $N$ databases given by,
    \begin{align}
        A_n^{[V(i)]}&=S_n^TQ_n^{[V(i)]}\\
        &=\frac{1}{f_1-\alpha_n}W_{\theta,1}^{[V(i)]}+\dotsc+\frac{1}{f_{\ell}-\alpha_n}W_{\theta,\ell}^{[V(i)]}\nonumber\\
        &\quad+P_{\alpha_n}(3\ell+1),
    \end{align}
    from which the $\ell$ bits of subpacket $V(i)$, $i\in\{1,\dotsc,|\tilde{V}|\}$ can be obtained from the $N$ answers, given that $N=\ell+3\ell+2=4\ell+2$ is satisfied. Thus, the subpacketization is $\ell=\frac{N-2}{4}$, and the reading cost is,
    \begin{align}
        \!\!C_R\!=\!\frac{P\log_qP\!\!+\!|\tilde{V}|(N\!\!+\!\log_q\!P)}{L}\!=\!\frac{4r'\!\!+\!\frac{4}{N}(1\!+\!r')\!\log_q \!P}{1-\frac{2}{N}},
    \end{align}
    where $r'$, $0\leq r'\leq 1$ is the sparsification rate in the downlink characterized by $|\tilde{V}|=P\times r'$.
\end{enumerate} 
   
\begin{remark}
The reading phase assumes that users at time $t$ download all subpackets updated at time $t-1$, which might result in a large $r'$. However, there exist methods such as \cite{GGS} that perform sparsification in the downlink as well (reduced $r'$). The proposed scheme is still applicable for any such setting and achieves the reading cost in \eqref{re} with the new $r'$.
\end{remark}

\subsection{Writing Phase at Time $t$}\label{write}

The writing phase of the PRUW scheme with sparsification consists of the following steps.
\begin{enumerate}
\item The user generates combined updates (one bit per subpacket) of the non-zero subpackets and has zero as the combined update of the rest of the $P(1-r)$ subpackets. The update of subpacket $s$ for database $n$ is given by,\footnote{A permuted version of these updates is sent to the databases.}
\begin{align}\label{update}
    \!U_n(s)\!\!=\!\!\begin{cases}
    0, & \text{$s\in B^c$},\\
    \sum_{i=1}^\ell\! \Tilde{\Delta}_{\theta,i}^{[s]}\! \prod_{j=1,j\neq i}^\ell (f_j\!-\!\alpha_n)\\
    \quad +\prod_{j=1}^\ell (f_j-\alpha_n)\hat{Z}_s, & \text{$s\in B$},
    \end{cases}
\end{align}
where $B$ is the set of subpacket indices with non-zero updates, $\hat{Z}_s$ is a random noise bit and $\Tilde{\Delta}_{\theta,i}^{[s]}=\frac{\Delta_{\theta,i}^{[s]}}{\prod_{j=1,j\neq i}^\ell (f_j-f_i)}$ with $\Delta_{\theta,i}^{[s]}$ being the update for the $i$th bit of subpacket $s$ of $W_\theta$. 
\item The user permutes the updates of subpackets using $\tilde{P}$,
\begin{align}
    \hat{U}_n(i)=U_n(\tilde{P}(i)),\quad i=1,\dotsc,P.
\end{align}
\item Then, the user sends the following (update, position) pairs to each database $n$,
\begin{align}
    Y_n^{[j]}=(\hat{U}_n^{[j]}, k^{[j]}), \quad j=1,\dotsc,Pr,
\end{align}
where $k^{[j]}$ is the $j$th non-zero subpacket index (permuted) based on $\tilde{P}$ and $\hat{U}_n^{[j]}$ is the corresponding update.
\item Based on the received (update, position) pairs, each database constructs an update vector $\tilde{U}_n$ of size $P\times 1$ with $\hat{U}_n^{[j]}$ placed as the $k^{[j]}$th entry and zeros elsewhere,
\begin{align}\label{permute}
    \tilde{U}_n=\sum_{j=1}^{Pr} \hat{U}_n^{[j]}\be_P(k^{[j]}).
\end{align}
\item $\tilde{U}_n$ in \eqref{permute} contains the combined updates of the form \eqref{update} arranged in a random permutation given by $\tilde{P}$. The databases are unable to determine the true indices of all zero subpackets since $\tilde{P}$ is not known by the databases. However, for correctness in writing phase, the updates in $\tilde{U}_n$ must be rearranged in the correct order. This is done with the permutation-reversing matrix given in \eqref{rearrange} as,
\begin{align}
    T_n&=R_n\tilde{U}_n=R\tilde{U}_n+\prod_{i=1}^\ell (f_i-\alpha_n)P_{\alpha_n}(\ell),
\end{align}
where $P_{\alpha_n}(\ell)$ is a $P\times1$ vector containing noise polynomials of $\alpha_n$ of degree $\ell$, $R\tilde{U}_n$ contains all updates of all subpackets (including zeros) in the correct order, while $\prod_{i=1}^\ell (f_i-\alpha_n)P_{\alpha_n}(\ell)$ contains random noise, that hides the indices of the zero update subpackets. 
\item The incremental update is calculated in the same way as described in \cite{ourICC} in each subpacket as,
\begin{align}
    h(s)\!\!&=D_n\times T_n(s)\times Q_n\\
    &=D_n\times U_n(s)\times Q_n+D_n\times P_{\alpha_n}(2\ell)\label{first}\\
    &=\begin{cases}
    \!\!\begin{bmatrix}
        \Delta_{\theta,1}^{[s]}\be_M(\theta) \\  \vdots\\ \Delta_{\theta,\ell}^{[s]} \be_M(\theta) 
    \end{bmatrix}\!\!+\!\!\begin{bmatrix}
    \!(f_1\!-\alpha_n)P_{\alpha_n}(2\ell)\!
        \\ \vdots\\
    \!(f_\ell\!-\alpha_n)P_{\alpha_n}(2\ell)\!
    \end{bmatrix}, & s\in B,\\
    \!\!\begin{bmatrix}
    (f_1-\alpha_n)P_{\alpha_n}(2\ell)
        \\ \vdots\\
    (f_{\ell}-\alpha_n)P_{\alpha_n}(2\ell)
    \end{bmatrix}, & s\in B^c
    \end{cases}\label{next}
\end{align}
where $P_{\alpha_n}(2\ell)$ here are noise vectors of size $M\ell\times1$ in \eqref{first} and $M\times1$ in \eqref{next} with polynomials of $\alpha_n$ of degree $2\ell$ and $D_n$ is the scaling matrix given by, 
\begin{align}
    D_n=\begin{bmatrix}
        (f_1-\alpha_n)I_M & \dotsc & 0\\
        \vdots & \vdots & \vdots\\
        0 & \dotsc & (f_\ell-\alpha_n)I_M\\
    \end{bmatrix}, 
\end{align}
for $ n\in\{1,\dotsc,N\}$. $h(s)$ is in the same format as the storage and hence can be added to the existing storage to obtain the updated storage, i.e.,
\begin{align}
    S_n^{[t]}(s)=S_n^{[t-1]}(s)+h(s), \quad s=1,\dotsc,P.
\end{align}
The writing cost of the scheme is given by,
\begin{align}
    C_W&= \frac{PrN(1+\log_q P)}{P\frac{N-2}{4}}=\frac{4r(1+\log_q P)}{1-\frac{2}{N}}.
\end{align}
\end{enumerate}

\begin{remark}
This problem can also be solved by considering a classical FSL setting (without sparsification) with $P$ submodels (i.e., $M=P$) and by using the private FSL scheme in \cite{ourICC} to update the sparse $Pr$ submodels. However, in this case the  normalized cost of sending the queries $Q_n$ given by $\frac{M\ell}{L}=\frac{P\ell}{L}=1$ is significantly large, and cannot be neglected.  
\end{remark}

\subsection{Example}
Assume that there are $P=5$ subpackets in each submodel. The coordinator first picks a random permutation of $\{1,\dotsc,5\}$ out of the $5!$ options available. Let the realization of the permutation be $\tilde{P}=\{2,5,1,3,4\}$. The corresponding permutation-reversing matrix for database $n$, $n\in\{1,\dotsc,N\}$, is given by,
\begin{align}\label{exR}
    R_n=\begin{bmatrix}
        0 & 0 & 1 & 0 & 0\\
        1 & 0 & 0 & 0 & 0\\
        0 & 0 & 0 & 1 & 0\\
        0 & 0 & 0 & 0 & 1\\
        0 & 1 & 0 & 0 & 0\\
    \end{bmatrix}+\prod_{i=1}^\ell(f_i-\alpha_n)\bar{Z},
\end{align}
where $\bar{Z}$ is a random noise matrix of size $5\times5$ and $\ell$ is the subpacketization. The coordinator places matrix $R_n$ at database $n$ at the beginning of the process and sends $\tilde{P}$ to each user. Assume that a given user wants to update submodel $\theta$ at time $t$. In the reading phase, the user only needs to download the subpackets that were updated at time $t-1$. Let the set of permuted indices of the subpackets updated by all users at time $t-1$ be $\tilde{V}=\{2,3\}$, which is known by all databases. One designated database sends these permuted indices to each of the users at time $t$ (who were also present at time $t-1$). Then, the user can obtain the real subpacket indices updated by all users at time $t-1$ using $V(i)=\tilde{P}(\tilde{V}(i))$ for $i=1,2$, i.e., $V=\{5,1\}$. The user sends the query specifying the requirement of submodel $\theta$ given by \eqref{query} to database $n$.\footnote{Note that the query vector is of size $M\ell\times1$ and is not considered in the cost calculation since $\frac{M\ell}{L}$ is negligible.} Then, the databases send the parameters of the two subpackets in $V$ as two bits, without learning the contents of $V$. To do this, each database first calculates the non-permuted query vector for each subpacket $V(i)$ using the permutation-reversing matrix and the query received. The query for subpacket $V(1)=5$ is,
\begin{align}\label{query_right}
    Q_n^{[5]}&=\begin{bmatrix}
        R_n(1,\tilde{V}(1))Q_n\\\vdots\\R_n(P,\tilde{V}(1))Q_n
    \end{bmatrix}=\begin{bmatrix}
        \mathbf{0}\\\mathbf{0}\\\mathbf{0}\\\mathbf{0}\\Q_n
    \end{bmatrix}+P_{\alpha_n}(\ell)
\end{align}
where $P_{\alpha_n}(\ell)$ is a vector of size $5M\ell\times1$ consisting of polynomials of $\alpha_n$ of degree $\ell$ and $\mathbf{0}$ is the all zeros vector of size $M\ell\times1$. Then, the answer from database $n$ corresponding to subpacket $V(1)=5$ is given by,
\begin{align}
   \! A_n^{[5]}&=S_n^TQ_n^{[5]}\\
    &=\frac{1}{f_1\!-\!\alpha_n}W_{\theta,1}^{[5]}\!+\dotsc+\!\frac{1}{f_{\ell}\!-\!\alpha_n}W_{\theta,\ell}^{[5]}\!+\!P_{\alpha_n}(3\ell\!+\!1),
\end{align}
from which the $\ell$ bits of subpacket $5$ of submodel $\theta$ can be correctly obtained if $N=4\ell+2$, which defines the subpacketization $\ell=\frac{N-2}{4}$. Similarly, the user can obtain subpacket 1 of $W_{\theta}$ by picking column $\tilde{V}(2)=3$ of $R_n$ in \eqref{query_right} and following the same process.

Once the user downloads and trains $W_{\theta}$, the user generates the $r$ fraction of subpackets with non-zero updates. Let the subpacket indices with non-zero updates be 1 and 4. The noisy updates generated by the user to be sent to database $n$ according to \eqref{update} is given by $U_n=[U_n(1),0,0,U_n(4),0]^T$ in the correct order. The user then permutes $U_n$ based on the given permutation $\tilde{P}$, i.e., $\hat{U}_n(i)=U_n(\tilde{P}(i))$ for $i=\{1,\dotsc,5\}$,
\begin{align}
    \hat{U}_n&=[0,0,U_n(1),0,U_n(4)]^T\label{perm}.
\end{align}
The user sends the values and the positions of the non-zero updates as $(U_n(1),3)$ and $(U_n(4),5)$ based on the permuted order. Each database receives these pairs and reconstructs \eqref{perm},
\begin{align}
    \tilde{U}_n&=U_n(1)\be_5(3)+U_n(4)\be_5(5)=\hat{U}_n.
\end{align}
To rearrange the updates back in the correct order privately, database $n$ multiplies $\tilde{U}_n$ by the permutation-reversing matrix,
\begin{align}
    T_n&=R_n\times\tilde{U}_n\\
    &=\begin{bmatrix}
        0 & 0 & 1 & 0 & 0\\
        1 & 0 & 0 & 0 & 0\\
        0 & 0 & 0 & 1 & 0\\
        0 & 0 & 0 & 0 & 1\\
        0 & 1 & 0 & 0 & 0\\
    \end{bmatrix}\tilde{U}_n+\prod_{i=1}^\ell(f_i-\alpha_n)\bar{Z}\times \tilde{U}_n\\
    &=[U_n(1),0,0,U_n(4),0]^T+\prod_{i=1}^\ell(f_i-\alpha_n)P_{\alpha_n}(\ell)\label{reorder},
\end{align}
since $U_n(1)$ and $U_n(4)$ are of the form $\sum_{i=1}^\ell \Tilde{\Delta}_{\theta,i} \prod_{j=1,j\neq i}^\ell (f_j-\alpha_n)+\prod_{j=1}^\ell (f_j-\alpha_n)\hat{Z}=P_{\alpha_n}(\ell)$. The incremental update of subpacket $s$, is calculated by,
\begin{align}
    h(s)&=D_n\times T_n(s)\times Q_n\\
    &=\begin{cases}
    \!\!\begin{bmatrix}
        \Delta_{1,1}^{[s]}\be_M(\theta) \\ \vdots\\ \Delta_{1,\ell}^{[s]} \be_M(\theta) 
    \end{bmatrix}+\begin{bmatrix}
        (f_1-\alpha_n)P_{\alpha_n}(2\ell)\\ \vdots\\
        (f_\ell-\alpha_n)P_{\alpha_n}(2\ell)
    \end{bmatrix}, & \!\!s=1,4\\
    \!\!\begin{bmatrix}
        (f_1-\alpha_n)P_{\alpha_n}(2\ell)\\ \vdots\\
        (f_\ell-\alpha_n)P_{\alpha_n}(2\ell)
    \end{bmatrix}, & \!\!s=2,3,5
    \end{cases}
\end{align}
where $P_{\alpha_n}(2\ell)$ are vectors of size $M\times1$ consisting of noise polynomials of $\alpha_n$ of degree $2\ell$. Since the incremental update is in the same format as the storage in \eqref{storage3}, the existing storage can be updated as $S_n^{[t]}(s)=S_n^{[t-1]}(s)+h(s)$ for $s=1,\dotsc,5$.

\newpage

\bibliographystyle{unsrt}
\bibliography{references.itw2022}

\end{document}